\documentclass[12pt,a4paper]{article}

\usepackage{graphicx}  

\usepackage{xspace}
\usepackage{color}
\usepackage{colortbl}

\usepackage{amsmath}

\usepackage{ifthen} 

\usepackage{mciteplus}

\newboolean{pdflatex}
\setboolean{pdflatex}{true} 
\newboolean{articletitles}
\setboolean{articletitles}{true} 

\newboolean{uprightparticles}
\setboolean{uprightparticles}{false} 
\usepackage{amssymb}
\usepackage{amsfonts}
\usepackage{upgreek}

\usepackage{hyperref}
\usepackage[all]{hypcap} 

\def\lhcb {LHCb\xspace}
\def\ux85 {UX85\xspace}

\def\lhc {LHC\xspace}

\def\babar  {BaBar\xspace}
\def\belle  {Belle\xspace}

\def\delphi {DELPHI\xspace}

\def\cdf    {CDF\xspace}

\def\velo   {VELO\xspace}

\def\hlt    {HLT\xspace}

\ifthenelse{\boolean{uprightparticles}}%
{

 \def\Ppi         {\ensuremath{\uppi}\xspace}

 \def\Ppsi        {\ensuremath{\uppsi}\xspace}

 \def\PDelta      {\ensuremath{\Delta}\xspace}                 
 \def\PXi      {\ensuremath{\Xi}\xspace}                 
 \def\PLambda      {\ensuremath{\Lambda}\xspace}                 
 \def\PSigma      {\ensuremath{\Sigma}\xspace}                 
 \def\POmega      {\ensuremath{\Omega}\xspace}                 
 \def\PUpsilon      {\ensuremath{\Upsilon}\xspace}                 
 
 \def\PB      {\ensuremath{\mathrm{B}}\xspace}                 
                  
 \def\PD      {\ensuremath{\mathrm{D}}\xspace}

 \def\PJ      {\ensuremath{\mathrm{J}}\xspace}                 
 \def\PK      {\ensuremath{\mathrm{K}}\xspace}

 \def\Pb      {\ensuremath{\mathrm{b}}\xspace}

 \def\Pi      {\ensuremath{\mathrm{i}}\xspace}

 \def\Ps      {\ensuremath{\mathrm{s}}\xspace}

}
{

 \def\Ppi         {\ensuremath{\pi}\xspace}

 \def\Ppsi        {\ensuremath{\psi}\xspace}                 
                  
 \mathchardef\PDelta="7101
 \mathchardef\PXi="7104
 \mathchardef\PLambda="7103
 \mathchardef\PSigma="7106
 \mathchardef\POmega="710A
 \mathchardef\PUpsilon="7107
                  
 \def\PB      {\ensuremath{B}\xspace}                 
                  
 \def\PD      {\ensuremath{D}\xspace}

 \def\PJ      {\ensuremath{J}\xspace}                 
 \def\PK      {\ensuremath{K}\xspace}

 \def\Pb      {\ensuremath{b}\xspace}

 \def\Pi      {\ensuremath{i}\xspace}

 \def\Ps      {\ensuremath{s}\xspace}

}

\def\squark    {\ensuremath{\Ps}\xspace}

\def\pion  {\ensuremath{\Ppi}\xspace}

\def\pip   {\ensuremath{\pion^+}\xspace}

\def\kaon  {\ensuremath{\PK}\xspace}
  \def\Kbar  {\kern 0.2em\overline{\kern -0.2em \PK}{}\xspace}

\def\Kz    {\ensuremath{\kaon^0}\xspace}
\def\Kzb   {\ensuremath{\Kbar^0}\xspace}
\def\KzKzb {\ensuremath{\Kz \kern -0.16em \Kzb}\xspace}
\def\Kp    {\ensuremath{\kaon^+}\xspace}
\def\Km    {\ensuremath{\kaon^-}\xspace}

\def\KpKm  {\ensuremath{\Kp \kern -0.16em \Km}\xspace}

  \def\Dbar    {\kern 0.2em\overline{\kern -0.2em \PD}{}\xspace}
\def\D       {\ensuremath{\PD}\xspace}

\def\Dz      {\ensuremath{\D^0}\xspace}
\def\Dzb     {\ensuremath{\Dbar^0}\xspace}
\def\DzDzb   {\ensuremath{\Dz {\kern -0.16em \Dzb}}\xspace}
\def\Dp      {\ensuremath{\D^+}\xspace}
\def\Dm      {\ensuremath{\D^-}\xspace}

\def\DpDm    {\ensuremath{\Dp {\kern -0.16em \Dm}}\xspace}

\def\Dstarp  {\ensuremath{\D^{*+}}\xspace}

\def\B       {\ensuremath{\PB}\xspace}
  \def\Bbar    {\kern 0.18em\overline{\kern -0.18em \PB}{}\xspace}

\def\Bs      {\ensuremath{\B^0_\squark}\xspace}

\def\jpsi     {\ensuremath{{\PJ\mskip -3mu/\mskip -2mu\Ppsi\mskip 2mu}}\xspace}

  \def\Y#1S{\ensuremath{\PUpsilon{(#1S)}}\xspace}

\newcommand{\decay}[2]{\ensuremath{#1\!\to #2}\xspace}         

\def\to                 {\ensuremath{\rightarrow}\xspace}

\def\order   {\ensuremath{\mathcal{O}}\xspace}

\def\CP                {\ensuremath{C\!P}\xspace}

\def\AT#1     {\ensuremath{A_{\mathrm{T}}^{#1}}\xspace}           

\def\C#1      {\ensuremath{\mathcal{C}_{#1}}\xspace}                       
\def\Cp#1     {\ensuremath{\mathcal{C}_{#1}^{'}}\xspace}                    
\def\Ceff#1   {\ensuremath{\mathcal{C}_{#1}^{\mathrm{(eff)}}}\xspace}        
\def\Cpeff#1  {\ensuremath{\mathcal{C}_{#1}^{'\mathrm{(eff)}}}\xspace}       
\def\Ope#1    {\ensuremath{\mathcal{O}_{#1}}\xspace}                       
\def\Opep#1   {\ensuremath{\mathcal{O}_{#1}^{'}}\xspace}                    

\def\ycp        {\ensuremath{y_{\CP}}\xspace}
\def\agamma     {\ensuremath{A_{\Gamma}}\xspace}

\newcommand{\tev}{\ensuremath{\mathrm{\,Te\kern -0.1em V}}\xspace}
\newcommand{\gev}{\ensuremath{\mathrm{\,Ge\kern -0.1em V}}\xspace}
\newcommand{\mev}{\ensuremath{\mathrm{\,Me\kern -0.1em V}}\xspace}
\newcommand{\kev}{\ensuremath{\mathrm{\,ke\kern -0.1em V}}\xspace}
\newcommand{\ev}{\ensuremath{\mathrm{\,e\kern -0.1em V}}\xspace}
\newcommand{\gevc}{\ensuremath{{\mathrm{\,Ge\kern -0.1em V\!/}c}}\xspace}
\newcommand{\mevc}{\ensuremath{{\mathrm{\,Me\kern -0.1em V\!/}c}}\xspace}
\newcommand{\gevcc}{\ensuremath{{\mathrm{\,Ge\kern -0.1em V\!/}c^2}}\xspace}
\newcommand{\gevgevcccc}{\ensuremath{{\mathrm{\,Ge\kern -0.1em V^2\!/}c^4}}\xspace}
\newcommand{\mevcc}{\ensuremath{{\mathrm{\,Me\kern -0.1em V\!/}c^2}}\xspace}

\def\mum  {\ensuremath{\,\upmu\rm m}\xspace}

\def\invpb {\ensuremath{\mbox{\,pb}^{-1}}\xspace}

\def\ps   {\ensuremath{{\rm \,ps}}\xspace}
\def\fs   {\ensuremath{\rm \,fs}\xspace}

\def\khz  {\ensuremath{{\rm \,kHz}}\xspace}

\def\order{{\ensuremath{\cal O}}\xspace}
\newcommand{\chisq}{\ensuremath{\chi^2}\xspace}

\def\gsim{{~\raise.15em\hbox{$>$}\kern-.85em
          \lower.35em\hbox{$\sim$}~}\xspace}
\def\lsim{{~\raise.15em\hbox{$<$}\kern-.85em
          \lower.35em\hbox{$\sim$}~}\xspace}

\def\tell1  {TELL1\xspace}
\def\ukl1   {UKL1\xspace}

\newcommand{\ie}{\mbox{\itshape i.e.}\xspace}

\def\deltam     {\ensuremath{\Delta m}\xspace}
\def\ipd        {\ensuremath{{\rm IP}_\PD}\xspace}
\def\ip         {\ensuremath{{\rm IP}}\xspace}

\def\dzkk       {\ensuremath{\decay{\Dz}{\Kp\Km}}\xspace}

\def\dzkpi      {\ensuremath{\decay{\Dz}{\Km\pip}}\xspace}

\def\SlowPi  {\ensuremath{\pion_{\rm s}}\xspace}

\def\Teslam  {\ensuremath{\mathrm{Tm}}\xspace}
\def\Am      {\ensuremath{\mathrm{A_{\rm m}}}\xspace}
\def\Ad      {\ensuremath{\mathrm{A_{\rm d}}}\xspace}

\begin{document}



\begin{titlepage}
\pagenumbering{roman}

\vspace*{-1.5cm}
\centerline{\large EUROPEAN ORGANIZATION FOR NUCLEAR RESEARCH (CERN)}
\vspace*{1.5cm}
\hspace*{-0.5cm}
\begin{tabular*}{\linewidth}{lc@{\extracolsep{\fill}}r}
\ifthenelse{\boolean{pdflatex}}
{\vspace*{-2.7cm}\mbox{\!\!\!\includegraphics[width=.14\textwidth]{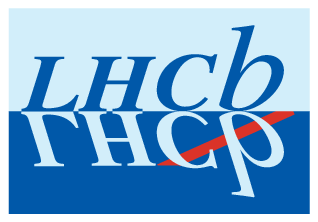}} & &}%
{\vspace*{-1.2cm}\mbox{\!\!\!\includegraphics[width=.12\textwidth]{figs/lhcb-logo}} & &}%
\\
 & & LHCb-PAPER-2011-032 \\  
 & & CERN-PH-EP-2011-206 \\  
 & & \today \\ 
\end{tabular*}

\vspace*{3.0cm}

{\bf\boldmath\huge
\begin{center}
Measurement of mixing and \CP violation parameters in two-body charm decays
\end{center}
}

\vspace*{1.5cm}

\begin{center}
The LHCb Collaboration
\footnote{Authors are listed on the following pages.}
\end{center}

\vspace{\fill}

\begin{abstract}
  \noindent
A study of mixing and indirect \CP violation in \Dz mesons through the determination of the parameters \ycp and \agamma is presented.
The parameter \ycp is the deviation from unity of the ratio of effective lifetimes measured in \Dz decays to the \CP eigenstate $\Kp\Km$ with respect to decays to the Cabibbo favoured mode $K^-\pi^+$. 
The result measured using data collected by \lhcb in 2010, corresponding to an integrated luminosity of $29\invpb$, is
\begin{displaymath}
\ycp = (5.5\pm6.3_{\rm stat}\pm4.1_{\rm syst})\times 10^{-3}.
\end{displaymath}
The parameter \agamma is the asymmetry of effective lifetimes measured in decays of \Dz and \Dzb mesons to $\Kp\Km$. 
The result is
\begin{displaymath}
\agamma = (-5.9\pm5.9_{\rm stat}\pm2.1_{\rm syst})\times 10^{-3}.
\end{displaymath}
A data-driven technique is used to correct for lifetime-biasing effects.
\end{abstract}

\vspace*{1.0cm}
\begin{center}
Submitted to JHEP
\end{center}
\vspace{\fill}

\end{titlepage}



\section*{The LHCb collaboration}
R.~Aaij$^{23}$, 
C.~Abellan~Beteta$^{35,n}$, 
B.~Adeva$^{36}$, 
M.~Adinolfi$^{42}$, 
C.~Adrover$^{6}$, 
A.~Affolder$^{48}$, 
Z.~Ajaltouni$^{5}$, 
J.~Albrecht$^{37}$, 
F.~Alessio$^{37}$, 
M.~Alexander$^{47}$
% ...

{\footnotesize \it \noindent
% ...
$ ^{5}$Clermont Universit\'{e}, Universit\'{e} Blaise Pascal, CNRS/IN2P3, LPC, Clermont-Ferrand, France\\
$ ^{6}$CPPM, Aix-Marseille Universit\'{e}, CNRS/IN2P3, Marseille, France\\
% ...
$ ^{23}$Nikhef National Institute for Subatomic Physics, Amsterdam, The Netherlands\\
% ...
$ ^{35}$Universitat de Barcelona, Barcelona, Spain\\
$ ^{36}$Universidad de Santiago de Compostela, Santiago de Compostela, Spain\\
$ ^{37}$European Organization for Nuclear Research (CERN), Geneva, Switzerland\\
% ...
$ ^{42}$H.H. Wills Physics Laboratory, University of Bristol, Bristol, United Kingdom\\
% ...
$ ^{47}$School of Physics and Astronomy, University of Glasgow, Glasgow, United Kingdom\\
$ ^{48}$Oliver Lodge Laboratory, University of Liverpool, Liverpool, United Kingdom\\
% ...
$ ^{n}$LIFAELS, La Salle, Universitat Ramon Llull, Barcelona, Spain\\
% ...
}

\cleardoublepage




\pagestyle{plain} 
\setcounter{page}{1}
\pagenumbering{arabic}


%

\section{Introduction}
\label{sec:introduction}

Mixing of neutral \Dz mesons has only recently been established~\cite{Aubert:2007wf,Staric:2007dt,Abe:2007rd} and first evidence for \CP violation in the charm sector has just been seen by \lhcb~\cite{Aaij:2011in}.
In this work the mixing and \CP violation parameters \ycp and \agamma in the decays of neutral \Dz mesons into two charged hadrons are studied.
Both quantities are measured here for the first time at a hadron collider.
The observable \ycp is the deviation from unity of the ratio of inverse effective lifetimes in the decay modes $\Dz\to\Kp\Km$ and $\Dz\to\Km\pip$
\begin{equation}
\ycp \equiv \frac{\hat{\Gamma}(\Dz\to\Kp\Km)}{\hat{\Gamma}(\Dz\to\Km\pip)}-1,
\label{eqn:ycp}
\end{equation}
where effective lifetime refers to the value measured using a single exponential model.
All decays implicitly include their charge conjugate modes, unless explicitly stated otherwise.
Similarly, \agamma is given by the asymmetry of inverse effective lifetimes as
\begin{equation}
\agamma \equiv \frac{\hat{\Gamma}(\Dz\to\Kp\Km)-\hat{\Gamma}(\Dzb\to\Kp\Km)}{\hat{\Gamma}(\Dz\to\Kp\Km)+\hat{\Gamma}(\Dzb\to\Kp\Km)}.
\label{eqn:agamma}
\end{equation}

The neutral \Dz mass eigenstates $|D_{1,2}\rangle$ with masses $m_{1,2}$ and widths $\Gamma_{1,2}$ can be expressed as linear combinations of the flavour eigenstates as $|D_{1,2}\rangle=p|\Dz\rangle\pm{}q|\Dzb\rangle$ with complex coefficients $p$ and $q$ satisfying $|p|^2+|q|^2=1$.
The average mass and width are defined as $m\equiv(m_1+m_2)/2$ and $\Gamma\equiv(\Gamma_1+\Gamma_2)/2$; the mass and width difference are used to define the mixing parameters $x\equiv(m_2-m_1)/\Gamma$ and $y\equiv(\Gamma_2-\Gamma_1)/(2\Gamma$).
The phase convention is chosen such that $\CP|\Dz\rangle=-|\Dzb\rangle$ and $\CP|\Dzb\rangle=-|\Dz\rangle$ which leads, in the case of no \CP violation ($p=q$), to $|D_{1}\rangle$ being the \CP odd and $|D_{2}\rangle$ the \CP even eigenstate, respectively.

The parameter
\begin{equation}
\lambda_f=\frac{q\bar{A}_f}{pA_f}=-\eta_{\CP}\left|\frac{q}{p}\right|\left|\frac{\bar{A}_f}{A_f}\right|e^{i\phi},
\end{equation}
contains the amplitude $A_f$ ($\bar{A}_f$) of \Dz (\Dzb) decays to the \CP eigenstate $f$ with eigenvalue $\eta_{\CP}$.
The mixing parameters $x$ and $y$ are known to be at the level of $10^{-2}$ while both the phase and the deviation of the magnitude from unity of $\lambda_f$ are experimentally only constrained to about $0.2$~\cite{Asner:2010qj}.
The direct \CP violation, \ie\ the difference in the rates of \Dz and \Dzb decays, is constrained to the level of $10^{-2}$ and has recently been measured by \lhcb~\cite{Aaij:2011in}.
Introducing $|q/p|^{\pm 2}\approx1\pm \Am$ and $|\bar{A}_f/A_f|^{\pm 2}\approx1\pm \Ad$, with the assumption that $\Am$ and $\Ad$ are small, and neglecting terms below $10^{-4}$ according to the experimental constraints, one obtains according to Ref.~\cite{Gersabeck:2011xj}
\begin{equation}
\ycp \approx \left( 1 +\frac{1}{8}\Am^2\right)y\cos\phi -\frac{1}{2}\Am x\sin\phi.
\label{eqn:ycptheory}
\end{equation}
In the limit of no \CP violation \ycp is equal to $y$ and hence becomes a pure mixing parameter.
However, once precise measurements of $y$ and \ycp are made, any difference between $y$ and \ycp would be a sign of \CP violation.

Previous measurements of \ycp have been performed by \babar and \belle.
The results are $\ycp=(11.6\pm2.2\pm1.8)\times 10^{-3}$~\cite{Aubert:2009ck} for \babar and $\ycp=(13.1\pm3.2\pm2.5)\times 10^{-3}$~\cite{Staric:2007dt} for \belle.
They are consistent with the world average of $y=(7.5\pm1.2)\times 10^{-3}$~\cite{Asner:2010qj}.

The study of the lifetime asymmetry of \Dz and \Dzb mesons decaying into the singly Cabibbo-suppressed final state $\Kp\Km$ can reveal indirect \CP violation in the charm sector.
The measurement can be expressed in terms of the quantity \agamma.
Using the same expansion as for \ycp leads to
\begin{eqnarray}
\agamma & \approx & \bigg[\frac{1}{2}(\Am+\Ad)y\cos\phi-x\sin\phi \bigg]\frac{1}{1+\ycp}\nonumber\\
& \approx & \frac{1}{2}(\Am+\Ad)y\cos\phi-x\sin\phi.
\label{eqn:agammatheory}
\end{eqnarray}
Despite this measurement being described in most literature as a determination of indirect \CP violation it is apparent that direct \CP violation at the level of $10^{-2}$ can have a contribution to \agamma at the level of $10^{-4}$.
Therefore precise measurements of both time-dependent and time-integrated asymmetries are necessary to reveal the nature of \CP violating effects in the \Dz system.

The measurement of \agamma requires tagging the flavour of the \Dz at production, which will be discussed in the following section.
Previous measurements of \agamma were performed by \belle and \babar leading to $\agamma=(0.1\pm3.0\pm1.5)\times 10^{-3}$~\cite{Staric:2007dt} and $\agamma=(2.6\pm3.6\pm0.8)\times 10^{-3}$~\cite{Aubert:2007en}, respectively.
They are consistent with zero, hence showing no indication of \CP violation.

\section{Data selection}
\label{sec:selections}

LHCb is a precision heavy flavour experiment which exploits the abundance of charm particles produced in collisions at the Large Hadron Collider (LHC).
The LHCb detector~\cite{Alves:2008zz} is a single arm spectrometer at the LHC with a pseudorapidity acceptance of $2<\eta<5$ for charged particles.  
High precision measurements of flight distances are provided by the Vertex Locator (\velo), which consists of two halves with a series of semi-circular silicon microstrip detectors.
The \velo measurements, together with momentum information from forward tracking stations and a $4~\Teslam$ dipole magnet, lead to decay-time resolutions of the order of one tenth of the \Dz lifetime.
Two Ring-Imaging Cherenkov (RICH) detectors using three different radiators provide excellent pion-kaon separation over the full momentum range of interest.
The detector is completed by hadronic and electromagnetic calorimeters and muon stations.
The measurements presented here are based on a data sample corresponding to an integrated luminosity of $29\invpb$ of $pp$ collisions at $\sqrt{s}=7\tev$ recorded during the LHC run in 2010.

\subsection{Trigger selection}
\label{sec:triggerdescription}
The \lhcb trigger consists of hardware and software (\hlt) stages. 
The hardware trigger is responsible for reducing the \lhc $pp$ interaction rate from \order{(10)}~MHz to
the rate at which the \lhcb subdetectors can be read out, nominally 1~MHz. 
It selects events based on the transverse momentum of track segments in the muon stations,
the transverse energy of clusters in the calorimeters, and overall event multiplicity.

The \hlt further reduced the event rate to about $2~\khz$ in 2010, at which the data was stored for offline processing. 
The \hlt runs the same software for the track reconstruction and event selection as is used offline and has access to the full event information. 

The first part of the \hlt is based on the reconstruction of tracks and primary interaction vertices in the \velo.
Heavy flavour decays are identified by their large lifetimes, which cause their daughter tracks to be
displaced from the primary interaction. The trigger first selects \velo tracks whose distance of closest approach
to any primary interaction, known as the impact parameter (IP), exceeds $110~\mum$. 
In addition the tracks are required to have at least ten hits in the VELO to reduce further the accepted rate of events.
This cut limits the fiducial volume for \Dz decays and therefore rejects events where the \Dz candidate has a large transverse component of the distance of flight,
causing an upper bound on the decay-time acceptance. 
The term decay-time acceptance will be used throughout this paper to refer to the selection efficiency as a function of the \Dz decay time. 
Selected tracks are then used to define a region of interest in the tracking stations after the dipole magnet,
whose size is defined by an assumed minimum track momentum of $8\gevc$; hits inside these search regions are used to form tracks traversing the full tracking system.
Tracks passing this selection are fitted,
yielding a full covariance matrix, and a final selection is made based on the track-fit quality and the track $\chisq(\ip)$.
The $\chisq(\ip)$ is a measure of the consistency with the hypothesis that the IP is equal to zero.
At least one good track is required for the event to be accepted.
The requirements on both the track IP and on the $\chisq(\ip)$ reduce the number of \Dz candidates with a short decay time.

In the second part of the \hlt, an exclusive selection of \Dz candidates is performed by reconstructing two-track vertices.
Further cuts are placed on the $\chisq(\ip)$ of the \Dz daughters and the displacement significance of the \Dz vertex from the primary interaction, as well
as a requirement which limits the collinearity angle between the \Dz momentum and the direction of flight, as defined by the primary and decay vertices.
These cuts all affect the distribution of the decay time of the \Dz candidates. Additional cuts are placed on track and vertex fit quality,
and on kinematic quantities such as the transverse momentum of the \Dz candidate, which have no effect on the decay-time distribution.

\subsection{Offline selection}
\label{sec:offline}

Given the abundance of charm decays, the selection has been designed to achieve high purity.
It uses similar requirements to those made in the trigger selection, though often with tighter thresholds.
In addition it makes use of the RICH information for separating kaons and pions to achieve a low misidentification rate.
A mass window of $\pm{}16\mevcc$ (about $\pm2\sigma$) is applied to the invariant mass of the two \Dz daughter particles using the appropriate mass hypotheses.
After these criteria have been applied there is negligible remaining cross-feed between the different two-body \Dz decay modes.

Flavour tagging of the \Dz decays is done by reconstructing the $\Dstarp\to\Dz\SlowPi^+$ decay, where the charge of the slow pion, \SlowPi, determines the flavour of the \Dz meson at production.
The selection applies loose requirements on the kinematics of the bachelor pion and the quality of the \Dstarp vertex fit.
The most powerful variable for selecting the \Dstarp decay is the difference in the reconstructed invariant masses of the \Dstarp and the \Dz candidates, \deltam.
Candidates are required to have \deltam in the range $|\deltam-145.4\mevcc|<2.0\mevcc$.

Events with multiple signal candidates are excluded from the analysis.
For tagged \Dz decays this causes a reduction of the number of candidates of about $15\%$ due to the high probability of assigning a random slow pion to form a \Dstarp candidate.
The numbers of selected candidates are 286,155 for \dzkpi and 39,262 for \dzkk decays.
\section{Determination of proper-time acceptance effects}
\label{sec:swimming}
Since absolute lifetime measurements are used to extract \ycp and \agamma, it is essential to correct for lifetime-biasing effects.
The analysis uses a data-driven approach that calculates, for each candidate and at every possible decay time, an acceptance value of zero or one which is related to the trigger decision and offline selection.
The final per-event acceptance function is used in the normalisation of the decay-time probability density function (PDF) as described in the following section.

The method used to determine decay-time acceptance effects is based on the so-called ``swimming'' algorithm.
This approach was first used at the NA11 spectrometer~\cite{Bailey:1985zz}, further developed within
\delphi\cite{Adam:1995mb} and \cdf \cite{Rademacker:2005ay,Aaltonen:2010ta}, studied at \lhcb~\cite{Gligorov:2008zza,Gersabeck:1210687}, and
applied to the measurement of the $\Bs\to\Kp\Km$ lifetime~\cite{LHCb-CONF-2011-018}.

The key to this method is the ability to execute the \lhcb trigger software, including the reconstruction, in precisely the same configuration used during data taking.
This is made possible by the implementation of all lifetime-biasing requirements of the trigger in software rather than in the hardware. 
The acceptance as a function of decay time is evaluated per event by artificially moving the position of the primary interaction vertices reconstructed in the trigger along the direction of the \Dz momentum in order to give the \Dz candidate a different decay time. 
In events containing multiple primary vertices, all are moved coherently in the direction of the \Dz momentum. 
An analogous procedure is used to obtain the decay-time acceptance of the offline selection.

\begin{figure}[htbp!!!!]
\centering
\begin{minipage}{0.32\textwidth}
(a)\\
\includegraphics[width=0.9\textwidth]{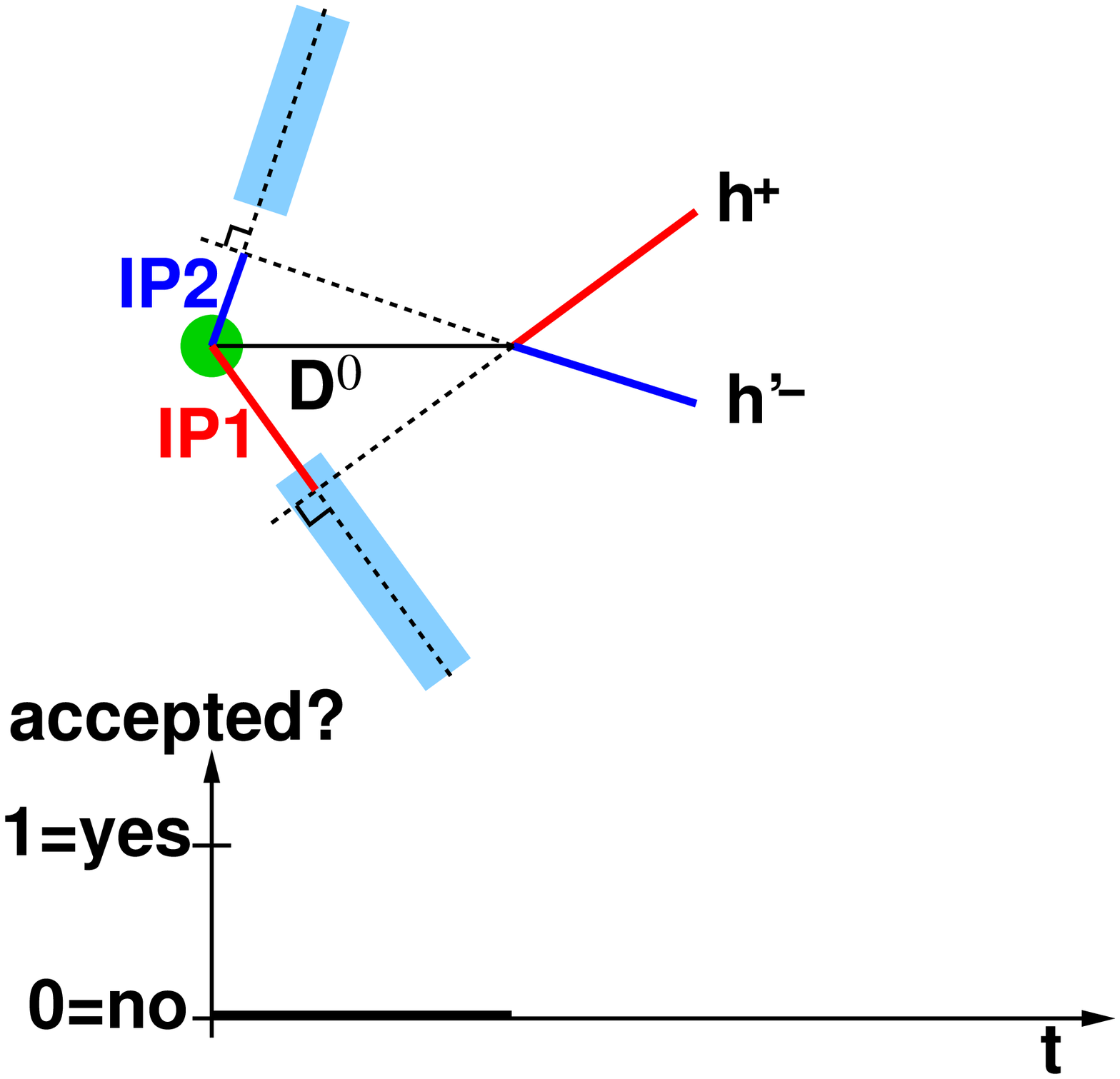}
\end{minipage}
\begin{minipage}{0.32\textwidth}
(b)\\
\includegraphics[width=0.9\textwidth]{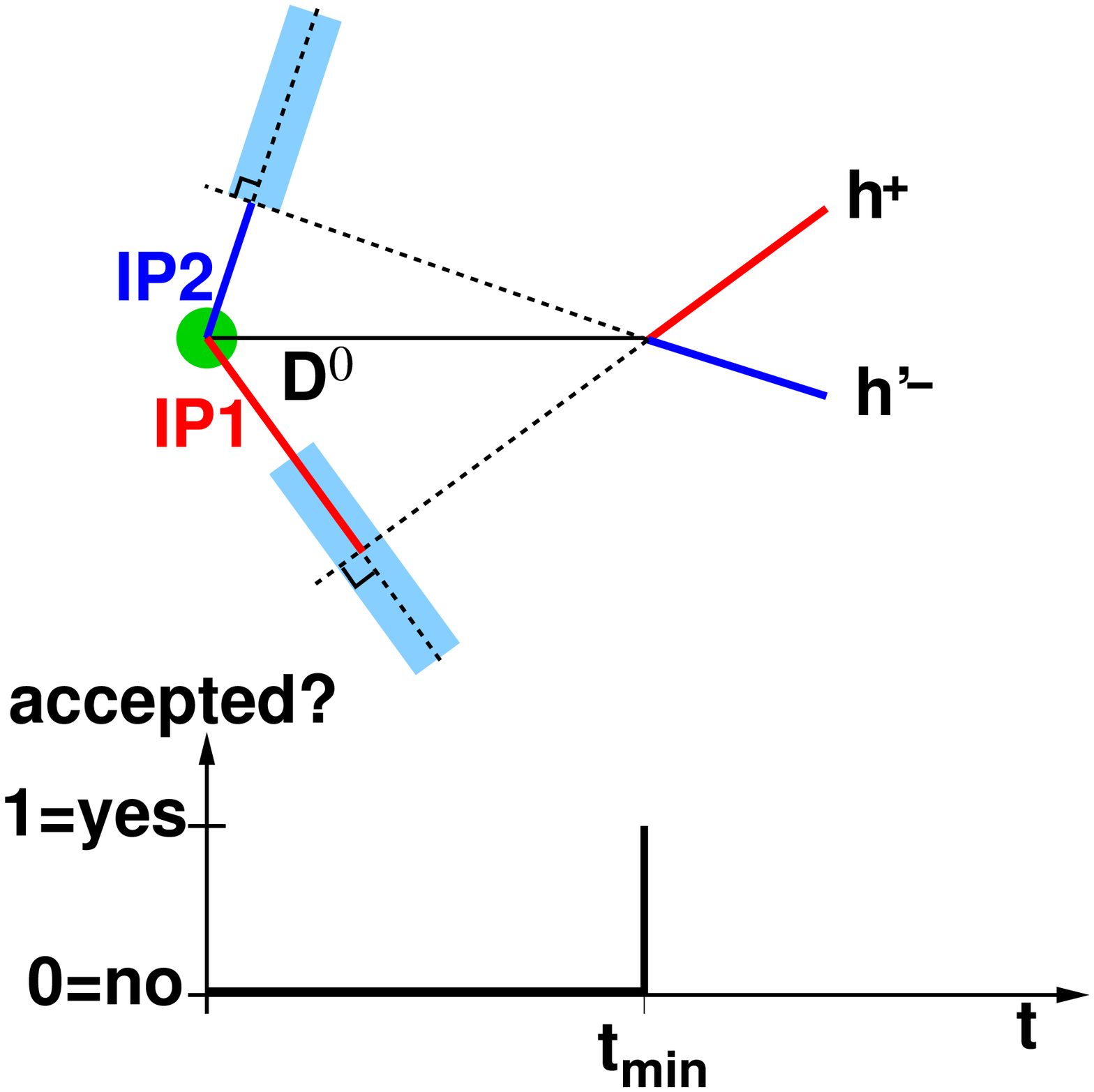}
\end{minipage}
\begin{minipage}{0.32\textwidth}
(c)\\
\includegraphics[width=0.9\textwidth]{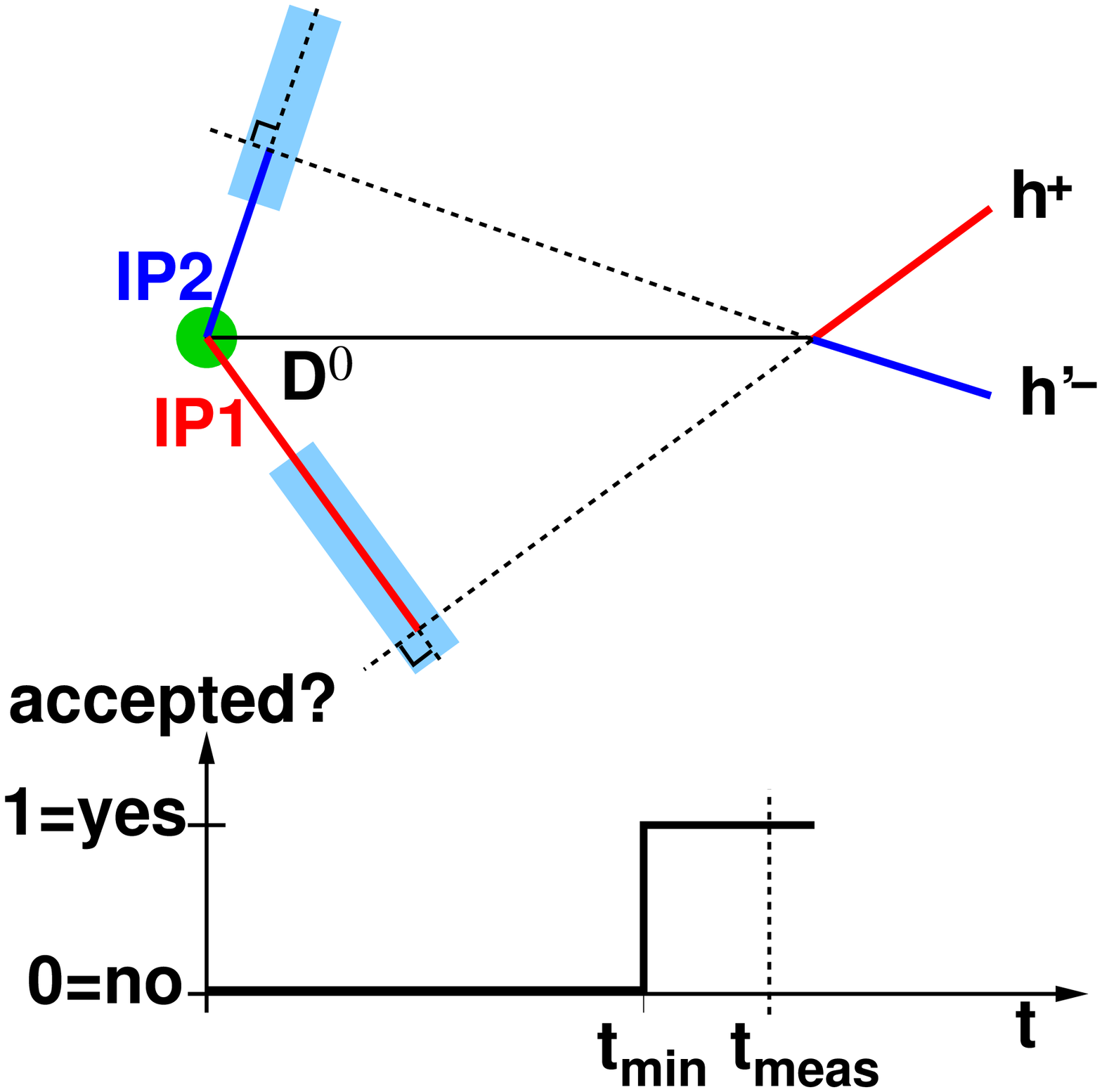}
\end{minipage}
\caption[Decay-time acceptance function for an event of a two-body hadronic decay.]{Variation of the decay-time acceptance function for a two-body \Dz decay when moving the primary vertex along the \Dz momentum vector. The shaded, light blue regions show the bands for accepting a track impact parameter. While the impact parameter of the negative track (IP2) is too low in (a) it reaches the accepted range in (b). The actual measured decay time, $t_{\rm meas}$, lies in the accepted region which continues to larger decay times (c).}
\label{fig:fit_swimming}
\end{figure}

A decay-time acceptance function for any single event is in the simplest case a step function, as shown in Fig.~\ref{fig:fit_swimming}, since the kinematics and chosen decay time of the \Dz decay fully determine whether the event is triggered by this \Dz candidate or not. 
It is important to note that the acceptance function for a given event does not depend on the measured decay time of that event, $t_{\rm meas}$.
Accepted (rejected) regions take an acceptance value of $1$ ($0$).
In this method decay-time independent selection efficiencies are factorised out and hence do not affect the result.
The presence of additional interaction vertices can lead to regions of no acceptance and the \velo geometry puts an upper limit on the accepted range.
Thus, a general decay-time acceptance function is given as a series of steps or top-hat functions.
The decay times at which the event enters or leaves one of these top hats are called turning points.
The acceptance functions of the trigger and offline selections are combined to a single acceptance function by including only the ranges which have been accepted by both selections.

\label{sec:velo_acc}
The idea of studying the decay-time dependence of the acceptance in principle requires moving the hits produced by the \Dz decay products.
The implementation of moving the primary vertices instead leads to significant technical simplifications.
However, this procedure ignores the fact that events are
no longer accepted if the mother particle has such a long decay time that one or both tracks can no longer be reconstructed inside the \velo.
This is a very small effect as a \Dz meson has to fly ten to a hundred times its average distance of flight in order to escape detection in the \velo.
Nevertheless, this effect can be estimated based on the knowledge of the position of the \velo modules and on the number of hits required to form a track.
Using the information on the position of the \velo sensors, the limit of the acceptance is determined by swimming the tracks along the \Dz momentum vector.
The result is treated as another per event decay-time acceptance and merged with the swimming results of the trigger and offline selections.

Finally, the track reconstruction efficiency in the trigger is reduced compared to the offline reconstruction due to the requirements described in Sect.~\ref{sec:selections}.
It has been verified, using a smaller sample acquired without a lifetime biasing selection, that this relative reconstruction efficiency does not depend on the decay time of the \Dz candidate with a precision of $3\times 10^{-3}$, and therefore introduces no significant additional acceptance effect.
\section{Fitting method}
\label{sec:methods}
\label{sec:method}

\begin{figure}
\centering
\includegraphics[width=0.49\textwidth]{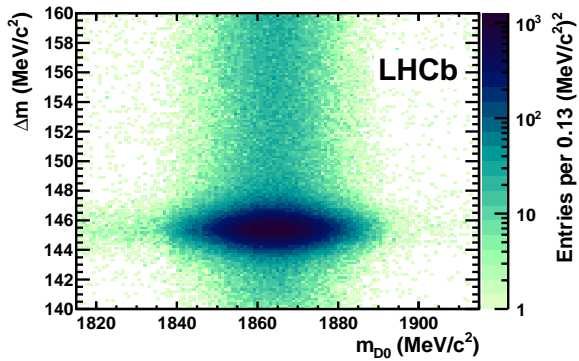}
\caption[]{\deltam vs $m_{\Dz}$ distribution for $\Dz\to\Km\pip$ candidates. The contribution of random slow pions extends around the signal peak in the vertical direction while background is visible as a horizontal band.}
\label{fig:md0_deltam}
\end{figure}

\begin{figure}[hbtp!!!!]
\centering
\includegraphics[width=0.49\textwidth]{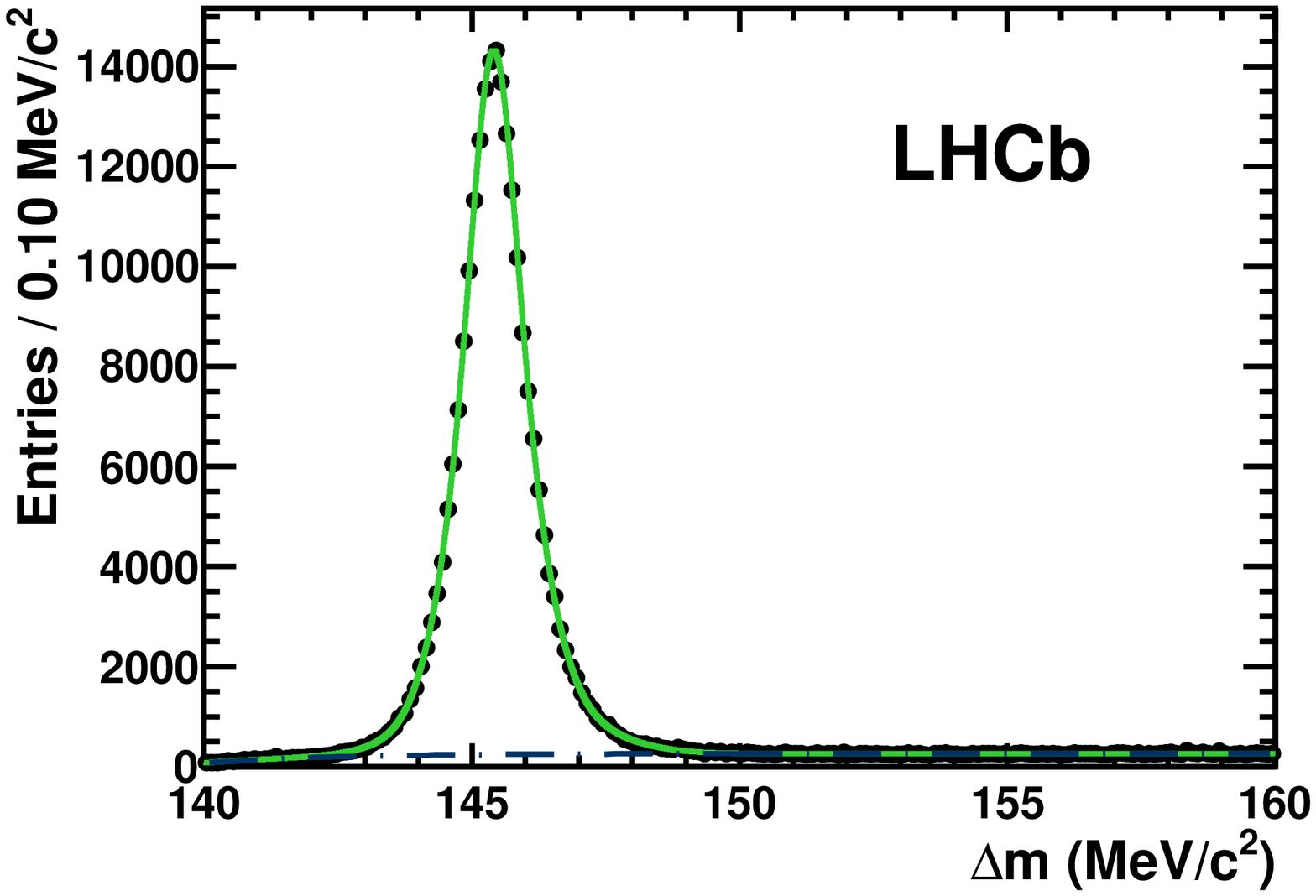}
\includegraphics[width=0.49\textwidth]{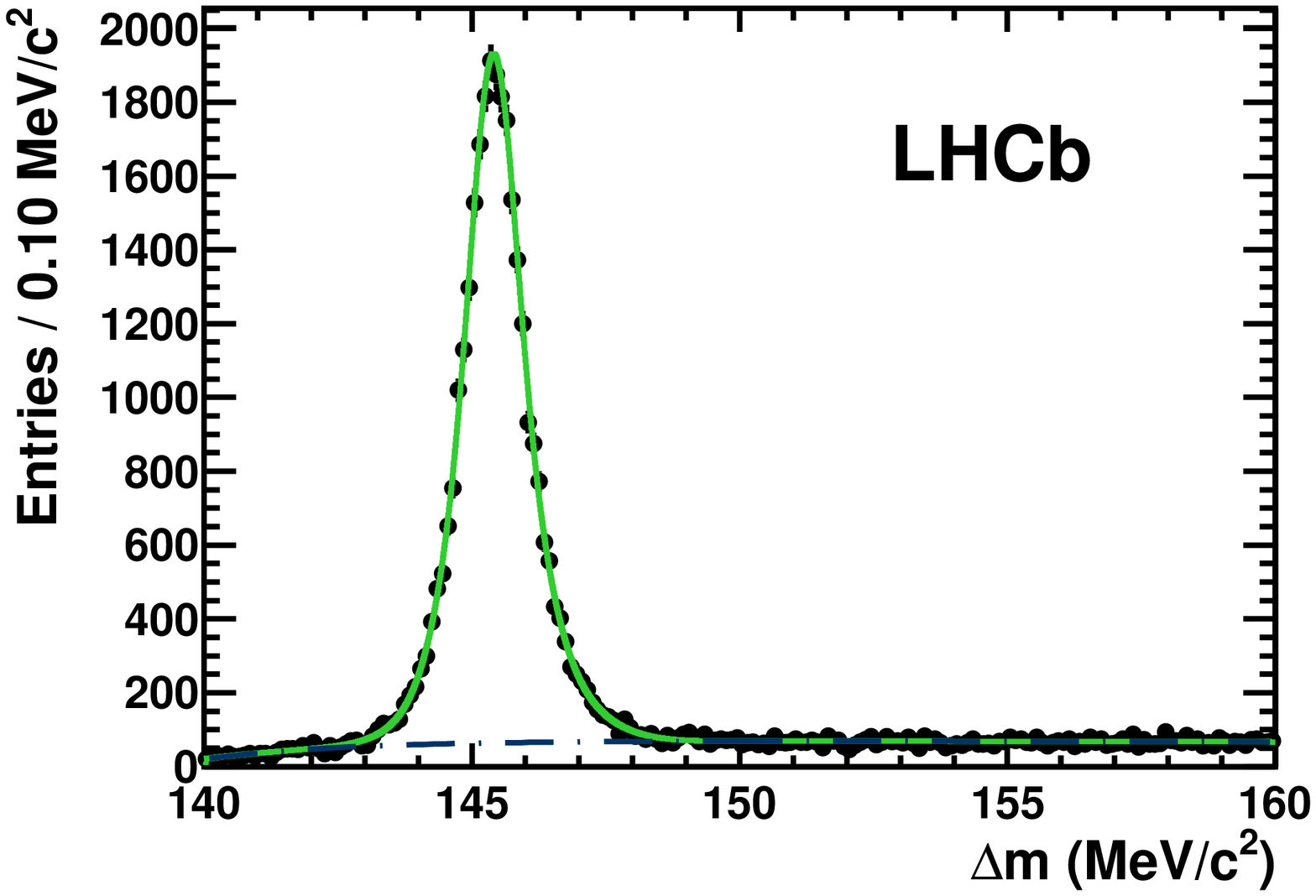}
\caption[]{\deltam fit projections of (left) $\Dz\to\Km\pip$ and (right) $\Dz\to\Kp\Km$ candidates. Shown are data (points), the total fit (green, solid) and the background component (blue, dot-dashed).}
\label{fig:result_tagged_d0kpi_deltam}
\end{figure}

The peak in \deltam from true \Dstarp decays is parametrised as the sum of three Gaussians; two of which have a common mean and a third which has a slightly higher mean.
The random \SlowPi background PDF is given by 
\begin{equation}
f_{\SlowPi}(\deltam) = 
\left( \frac{\deltam}{a} \right)^2 \: \left( 1-  \exp(-\frac{\deltam-d}{c}) \right) \: + \: b\left( \frac{\deltam}{d}-1 \right) \qquad\deltam\ge d,
\label{eqn:yield_roodstd0bg}
\end{equation}
where $a$ and $b$ define the slope at high values of \deltam, $c$ defines the curvature at low values of \deltam and $\deltam=d$ defines the threshold below which the function is equal to zero.
Figure~\ref{fig:md0_deltam} shows the \deltam vs $m_{\Dz}$ distribution and Fig.~\ref{fig:result_tagged_d0kpi_deltam} shows the fit to the mass difference between the reconstructed invariant masses of \Dstarp and \Dz candidates, \deltam.

%
%
The signal yield is extracted from fits to the reconstructed \Dz invariant mass distribution after application of the cut in \deltam.
The fit model for the signal peak has been chosen to be a double Gaussian and background is modelled as a first-order polynomial.
The background level is evaluated to be about $1\%$ for \dzkpi decays and about $3\%$ for \dzkk decays.
It consists of combinatorial background and partially reconstructed or misidentified \Dz decays.
If the latter stem from a \Dstarp decay they have a peaking distribution in \deltam similar to signal candidates.
The data in the mass sidebands are insufficient to reliably describe the background shape in other variables, so the background contribution is neglected in the time-dependent fit and a systematic uncertainty is estimated accordingly.

Events inside the signal windows in \deltam and $m_{D^0}$ are used in the lifetime fit, where \Dz mesons produced at the primary vertex (prompt) have to be distinguished from those originating from \Pb hadron decays (secondary).
The combined PDF for this decay-time dependent fit is factorized as
\begin{equation}
f(\chisq(\ipd),t,A)=\sum_{\substack{{\rm class}\\={\rm prompt},\\{\rm secondary}}} f_\text{IP}(\chisq(\ipd)|t,A,{\rm class})\: f_t(t|A,{\rm class}) \: f_\text{TP}(A|{\rm class})\: P({\rm class}).
\label{eqn:fit_time_complete}
\end{equation}
The four factors on the right-hand side of Eq.~\ref{eqn:fit_time_complete}, which will be described in detail below, are:
\begin{itemize}
\item the time-dependent PDFs for the $\ln\chisq(\ipd)$ values for prompt and secondary \Dz mesons;
\item the decay-time PDFs for prompt and secondary \Dz mesons;
\item the PDF for the turning points which define the acceptance $A$;
\item the fractions of prompt and secondary \Dz decays among the signal candidates.
\end{itemize}

The separation of prompt and secondary \Dz mesons is done on a statistical basis using the impact parameter of the \Dz candidate with respect to the primary vertex, \ipd.
For prompt decays, this is zero up to resolution effects, but can acquire larger values for secondary decays as the \Dz candidate does not in general point back to the primary vertex.
Given an estimate of the vertex resolution is available on an event-by-event basis, it is advantageous to use the \chisq of the \ipd instead of the impact parameter value itself.
The natural logarithm of this quantity, $\ln(\chisq(\ipd))$, allows for an easier parametrisation.
Empirically, the sum of two bifurcated Gaussians, \ie\ Gaussians with different widths on each side of the mean, and a third, symmetric Gaussian, all sharing a common peak position, is found to be a suitable model to describe the $\ln(\chisq(\ipd))$ distribution for both prompt and secondary \Dz.

For the prompt \Dz class the $\ln(\chisq(\ipd))$ distribution does not change with \Dz decay time as the true value is zero at all times and the resolution of $\ipd$ can be assumed to be independent of the measured decay time.
For secondary \Dz decays the decay-time and $\ln(\chisq(\ipd))$ are correlated.
The width of the $\ln(\chisq(\ipd))$ distribution is found to be approximately constant in decay time for both prompt and secondary \Dz mesons.
As Monte Carlo simulation studies suggest that secondary decays have a larger width in this variable, a scale factor between the widths for prompt and secondary mesons is introduced.
The mean value of $\ln(\chisq(\ipd))$ increases with \Dz decay time, which reflects the fact that \Dz mesons coming from other long-lived decays do not necessarily point back to the primary vertex and that they may point further away the further their parent particle flies.
The functional form for this time dependence is based on simulation and all parameters are determined in the fit to data.

The decay-time PDF, $f_t(t|A,{\rm class})$ is modelled as a single exponential for the prompt \Dz class and as a convolution of two exponentials for secondary decays.
To account for resolution effects, these are convolved with a single Gaussian resolution function.
The parameters of the resolution model are obtained from a fit to the decay time distribution of prompt \jpsi events. 
The resulting dilution is equivalent to that of a single Gaussian with a width of $50\fs$~\cite{Aaij:2011ji}.
The decay-time probability densities are properly normalized by integrating their product with the acceptance function $A$, evaluated by the swimming method, only over the decay-time intervals for which the event would have been accepted.
Hence, the acceptance turning points are used as boundaries in the integration.

Finally, a PDF for the per-event acceptance function is needed.
While the first acceptance turning point, \ie\ the one with the smallest decay time, depends on the \Dz decay topology, the others are governed more by the underlying event structure, e.g.\ the distribution of primary vertices.
The primary vertex distribution is independent of whether the \Dz candidate is of prompt or secondary origin.
Hence, the PDF can be approximated as $f_\text{TP}(A|{\rm class})\approx f_\text{TP}(\text{TP}_1|{\rm class})$, where $\text{TP}_1$ denotes the position of the first turning point.
The distribution for $f_\text{TP}(\text{TP}_1|{\rm prompt})$ is obtained by applying a cut at $\ln\chi^2(\ipd)<1$, thus selecting a very pure sample of prompt decays.
The distribution for $f_\text{TP}(\text{TP}_1|{\rm secondary})$ is obtained from the distribution of $\text{TP}_1$ weighted by the probability of each candidate being of secondary decay origin.

An initial fit is performed using the full $\ln\chisq(\ipd)$ distribution and all parameters in the description of this term are then fixed in the final fit.
A cut is then applied requiring $\ln\chisq(\ipd)<2$ in order to suppress the fractions of both background and secondary candidates to less than a few percent.
The final fit is performed on this reduced sample.
The effect of this procedure is estimated in the systematic uncertainty evaluation.

\section{Cross-checks and systematic uncertainties}
\label{sec:systematics}

The method for absolute lifetime measurements described in Sect.~\ref{sec:methods} comprises three main parts whose accuracy and potential for biasing the measurement have to be evaluated in detail:
\begin{itemize}
\item the determination of the event-by-event decay-time acceptance; 
\item the separation of prompt from secondary charm decays;
\item the estimation of the decay time distribution of combinatorial background.
\end{itemize}
Since the contribution of combinatorial background is ignored in the fit, it is important to evaluate the corresponding systematic uncertainty.
Furthermore, several other parameters are used in the fit whose systematic effects have to be evaluated, e.g.\ the description of the decay-time resolution.
It is generally expected that the systematic uncertainties in \ycp are similar to or larger than those in \agamma as in \ycp two different final states contribute to the measurement.

Several consistency checks are performed by splitting the dataset into subsets.
The stability is tested as a function of run period, \Dz momentum and transverse momentum, and primary vertex multiplicity.
No significant trend is observed and therefore no systematic uncertainty assigned.

The fitting procedure is verified using simplified Monte Carlo simulation studies.
No indication of a bias is observed and the statistical uncertainties are estimated accurately.

As an additional check, a control measurement is performed using the lifetime asymmetry of \Dz and \Dzb decays to the Cabibbo favoured decay \dzkpi.
The result is in agreement with zero and the flavour-averaged \Dz lifetime is found to be consistent with the world average.
Detailed results are given in Sect.~\ref{sec:results}.
The fit results for \dzkk decays were not looked at throughout the development of the method and the study of systematic uncertainties for the analyses of \ycp and \agamma.

\subsection{Evaluation of systematic uncertainties}

\label{sec:lengthScale}
Particle decay times are measured from the distance between the primary vertex and secondary decay vertex in the \velo.  
The systematic uncertainty from the distance scale is determined by considering the potential error on the length scale of the detector from the mechanical survey, thermal expansion and the current alignment precision.
A relative systematic uncertainty of $0.1\%$ is assigned to the measurements of absolute lifetimes, translating into a relative uncertainty of $0.1\%$ on \agamma and \ycp.

The method to evaluate the turning points of the decay-time acceptance functions described in Sect.~\ref{sec:swimming} uses an iterative approach which estimates the turning points to a precision
of about $1\fs$.
Two scenarios have been tested: a common bias of all acceptance turning points and a common length scaling of the turning points, which could originate from differences in the length scale in the trigger and offline reconstructions.
From a variation of the bias and the scale, a systematic uncertainty of $0.1\times 10^{-3}$ on \agamma and \ycp is determined.

\label{s:recobiases}
The reconstruction acceptance is dominated by the \velo geometry, which is accounted for by the method described in Sect.~\ref{sec:velo_acc}.
This leads to a correction of less than $1\fs$ on the absolute lifetime measurements, \ie\ a relative correction of about $0.24\%$.
No further systematic uncertainty is assigned to \agamma or \ycp as the size of this relative correction is negligible.
Additional studies of the reconstruction efficiency as a function of variables governing the decay geometry did not provide any indication of lifetime biasing effects.

\label{sec:sys_t_resolution}
The decay-time resolution is modelled by a single Gaussian.
The width of the resolution function is varied from its nominal value of $0.05\ps$ between $0.03\ps$ and $0.07\ps$.
The range of variation was chosen to cover possible alignment effects as well as effects from the different final state used to evaluate the resolution.
The result leads to a systematic uncertainty of $0.1\times 10^{-3}$ for \agamma and \ycp.

The fit range in decay time is restricted by lower and upper limits.
The lower limit is put in place to avoid instabilities in regions with extremely low decay-time acceptances and very few events.
The default cut value is $0.25\ps$ which is close to the lower end of the observed range of events.
This cut is varied to both $0.2\ps$ and $0.3\ps$.
The result leads to a systematic uncertainty of $0.1\times 10^{-3}$ for \agamma and $0.8\times 10^{-3}$ for \ycp.

The upper limit of the fit range in decay time is put in place to minimise the impact of long-lived background events.
The default cut is put at $6\ps$ which corresponds to about $15$ \Dz lifetimes.
This cut is varied to $5\ps$ and $8\ps$.
The result  leads to a systematic uncertainty of $0.2\times 10^{-3}$ for \agamma and \ycp.

The description of the contribution from combinatorial background is studied by varying its relative amount in the data sample and repeating the fit.
This is done by changing the \deltam window from the default of $\pm2\mevcc$ to $\pm1\mevcc$ and $\pm3\mevcc$. 
The result leads to a systematic uncertainty of $1.3\times 10^{-3}$ for \agamma and $0.8\times 10^{-3}$ for \ycp.

Events that originate from secondary charm decays are the background with the largest impact on the fit procedure as they have a very different decay-time distribution compared to prompt charm decays, but they peak in the invariant mass and \deltam distributions.
Also a fraction of combinatorial background events appear to be secondary-like in their $\ln\chisq(\ipd)$ distribution.
The cut of $\ln\chisq(\ipd)<2$ removes a large fraction of secondary-like events.
However, it is important that the remainder is properly modelled and does not bias the signal lifetime.
Varying this cut changes the relative number of secondary-like decays in the sample and therefore tests the stability of the secondary description in the fit model.
The fraction of secondary-like combinatorial background events is also altered with this test.
The $\ln\chisq(\ipd)$ cut is varied from $1.5$ which is just above the peak of the prompt distribution to $3.5$ where the probability densities for prompt and secondary decays are about equal.
The result leads to a systematic uncertainty of $1.6\times 10^{-3}$ for \agamma and $3.9\times 10^{-3}$ for \ycp.
The uncertainty is significantly larger for \ycp than for \agamma as may be expected from the difference in the background level in the channels involved in the \ycp measurement.

Additional studies were performed to estimate the potential impact of neglecting background events in the fit.
A background component was added to a simplified simulation.
The background decay time distribution was generated using extreme values of fits to the distribution observed in mass sidebands.
The average bias on the measurement of \ycp was about $2\times 10^{-3}$.
Since this is consistent with the assigned systematic uncertainty, we do not assign any additional uncertainty.

Furthermore, a background component was added to the \Dz decay-time PDF with a fixed fraction and average lifetime.
The fraction of this component, which was assumed to be secondary-like, was varied.
A change in the fit result for \ycp of $0$ (all background secondary-like) to $4\times 10^{-3}$ (all background prompt-like) was observed.
As it is known that a fraction of the background events are secondary-like, this result is considered consistent with the simplified simulation results.

\subsection{Summary of systematic uncertainties}

Table~\ref{tab:sys_summary} summarises the systematic uncertainties evaluated as described above.
The main systematic uncertainties are due to neglecting the combinatorial background and to the contribution of secondary-like decays. 
The total systematic uncertainties for \agamma and \ycp, obtained by combining all sources in quadrature, are $2.1\times 10^{-3}$ and $4.1\times 10^{-3}$, respectively.

\begin{table}
\caption[Systematic uncertainties.]
{Summary of systematic uncertainties.}
\vspace{2mm}
\centering
\begin{tabular}{|l|c|c|} 
\hline
Effect                                & \agamma $(10^{-3})$ & \ycp $(10^{-3})$\\
\hline
Decay-time acceptance correction                  & $0.1$ & $0.1$\\
Decay-time resolution                           & $0.1$ & $0.1$\\
Minimum decay-time cut                            & $0.1$ & $0.8$\\
Maximum decay-time cut                           & $0.2$ & $0.2$\\
Combinatorial background                    & $1.3$ & $0.8$\\
Secondary-like background                      & $1.6$ & $3.9$\\
\hline
Total                                             & $2.1$ & $4.1$\\
\hline
\end{tabular}
\label{tab:sys_summary}
\end{table}

\section{Results and conclusion}
\label{sec:results}
\begin{figure}[hbtp!!!!]
\centering
\includegraphics[width=0.49\textwidth]{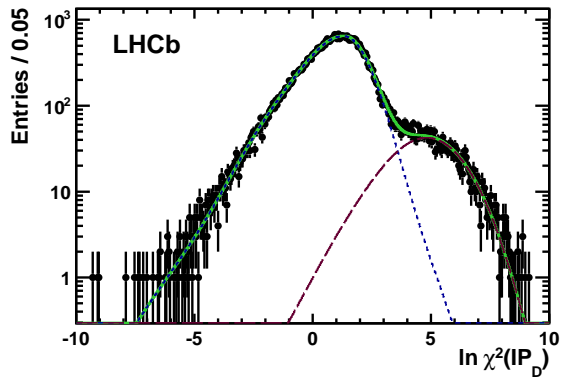}
\caption[]{$\ln\chi^2(\ipd)$ fit projection of $\Dz\to\Kp\Km$ candidates in logarithmic scale. Shown are data (points), the total fit (green, solid), the prompt signal (blue, short-dashed), and the secondary signal (purple, long-dashed).}
\label{fig:result_time_tagged_lnIPchi2_KK}
\end{figure}

The measurement of \ycp is based on absolute lifetime measurements as described in Sect.~\ref{sec:methods}.
It uses flavour-tagged events reconstructed in the decay chain $\Dstarp\to\Dz\pip$, with \Dz and \Dzb decays fitted simultaneously per decay mode.
The $\ln\chi^2(\ipd)$ projection of the final fit is shown in Fig.~\ref{fig:result_time_tagged_lnIPchi2_KK}.

The result for the lifetime measured in \dzkpi decays is $\tau(\Dz) = 410.2\pm0.9\fs$ where the uncertainty is statistical only.
The result for the lifetime is found to be in agreement with the current world average~\cite{Nakamura:2010zzi}.
Combining with the \dzkk lifetime measurement, $\tau(\Dz) = 408.0\pm2.4_{\rm stat}\fs$, this leads to the final result for \ycp of
\begin{displaymath}
\ycp = (5.5\pm6.3_{\rm stat}\pm4.1_{\rm syst})\times 10^{-3}.
\label{eqn:ycpresult}
\end{displaymath}

The measurement of \agamma is performed based on the same dataset and applying the same fitting method as used for the extraction of \ycp.
A control measurement is performed using decays to the Cabibbo favoured mode \dzkpi by forming a lifetime asymmetry analogous to Eq.~\ref{eqn:agamma}.
The measured flavour-tagged lifetimes are effective parameters since the fitted distributions also include mistagged events.
For the control measurement using \dzkpi decays this contamination is ignored as it is negligible due to the Cabibbo suppression of the mistagged decays.
The result for the asymmetry is $\agamma^{\PK\Ppi,\mathrm{eff}} = (-0.9\pm2.2_{\rm stat})\times 10^{-3}$ which is consistent with zero, according to expectations.

\label{sec:agamma}
For the extraction of \agamma, the mistagged decays are taken into account by expressing the measured effective lifetimes, $\tau^\mathrm{eff}$, in terms of the flavour-tagged lifetimes, $\tau(\Dz)$ and $\tau(\Dzb)$, and the mistag rate, $\epsilon_\pm$, where the sign is according to the sign of the tagging pion:
\begin{eqnarray}
\tau^\mathrm{eff}(\Dz)  & \approx & ( 1 - \epsilon_+ ) \: \tau(\Dz) + \epsilon_+ \: \tau(\Dzb)\\
\tau^\mathrm{eff}(\Dzb) & \approx & ( 1 - \epsilon_- ) \: \tau(\Dzb) + \epsilon_- \: \tau(\Dz).
\end{eqnarray}
The mistag rates are assumed to be independent of the final state and are extracted from the favoured \dzkpi decays as half the fraction of the random slow pion background in the signal region of the \deltam distribution.
They are found to be about $1.8\%$.
The systematic uncertainty due to this correction is negligible.

\begin{figure}
\centering
\includegraphics[width=0.45\textwidth]{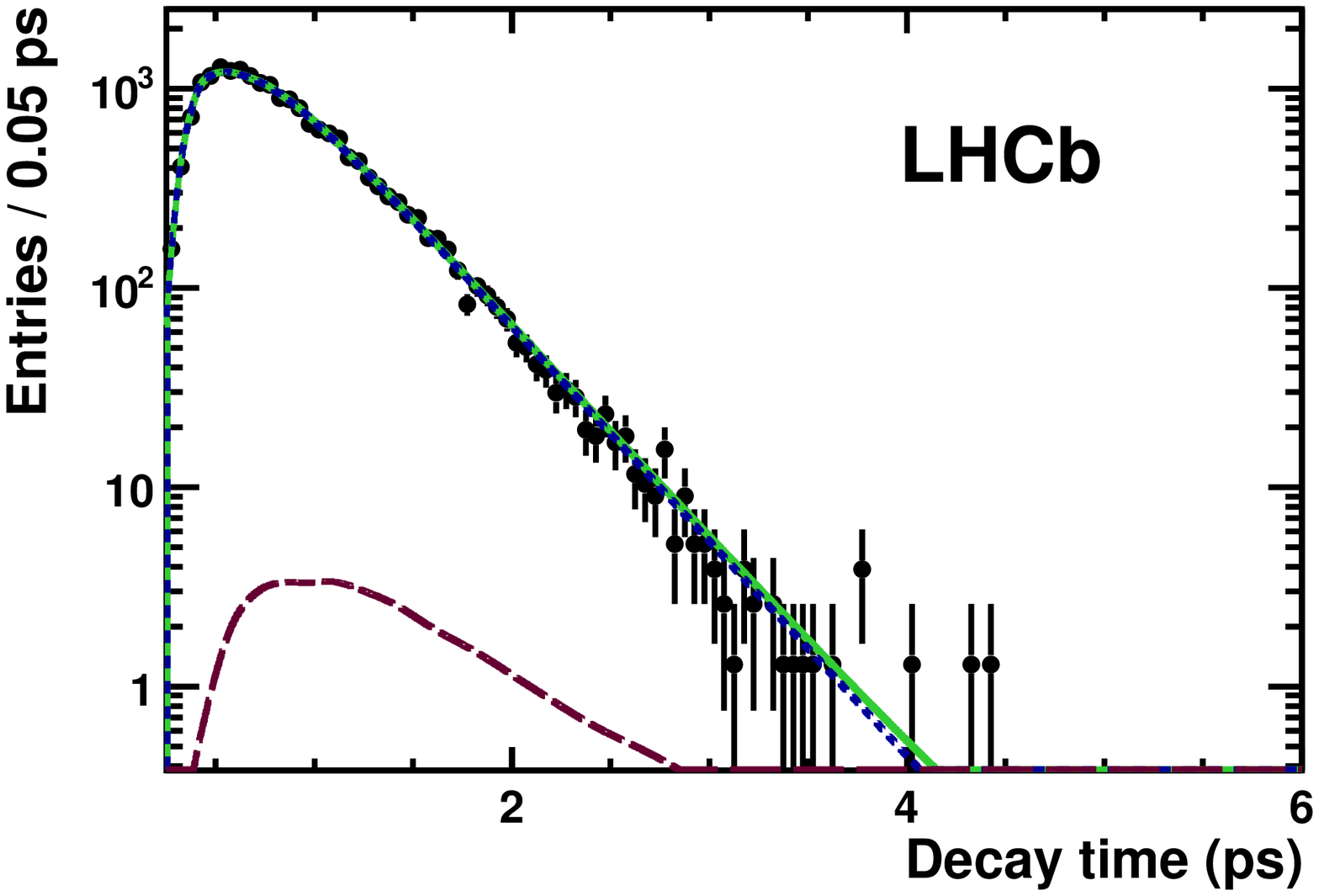}
\includegraphics[width=0.45\textwidth]{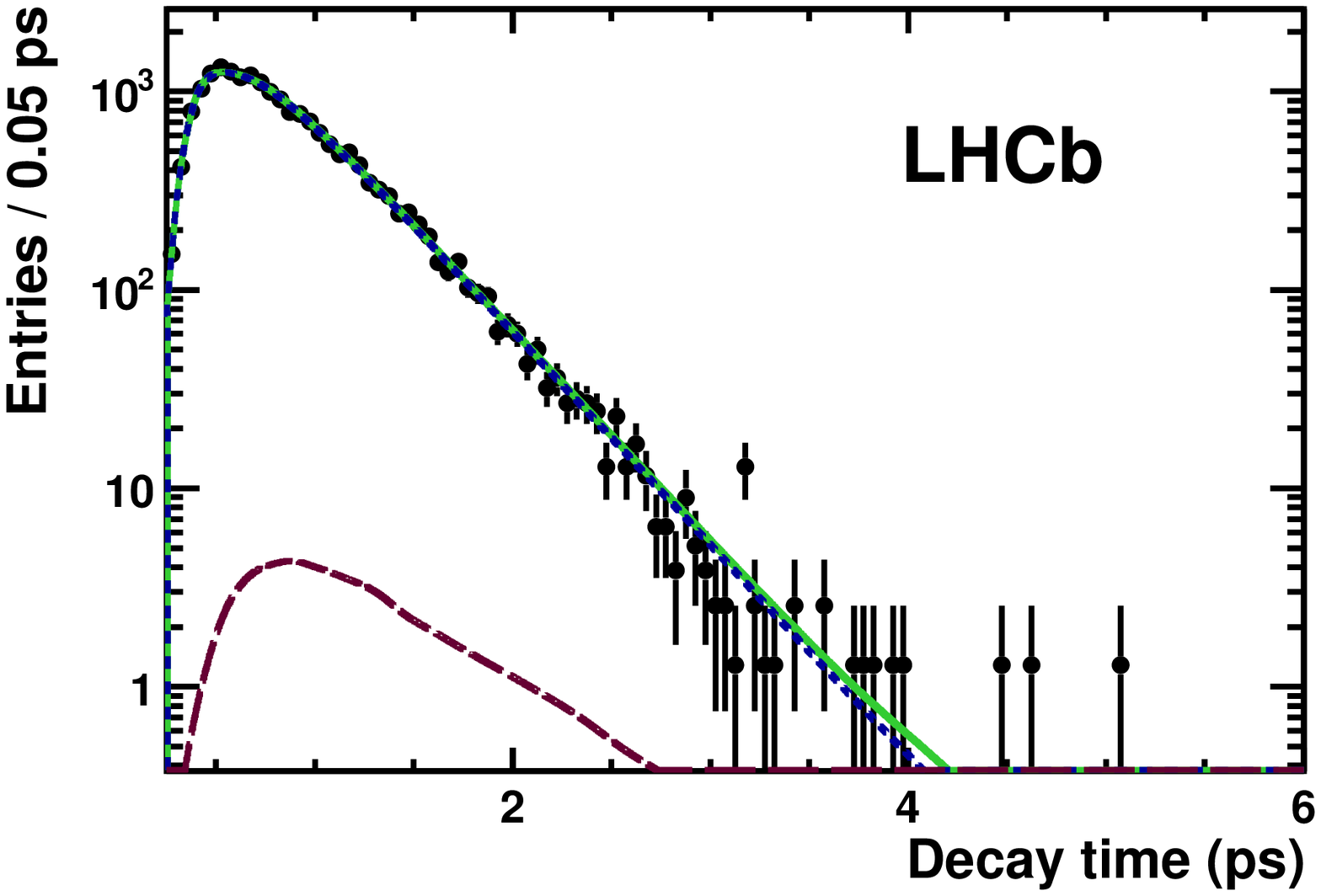}
\caption[]{Proper-time fit projections of (left) $\Dz\to\Kp\Km$ and (right) $\Dzb\to\Kp\Km$ candidates after application of the $\ln\chi^2(\ipd)<2$ cut. Shown are data (points),  the total fit (green, solid), the prompt signal (blue, short-dashed), and the secondary signal (purple, long-dashed).}
\label{fig:result_time_tagged_tau_KK}
\end{figure}
The projection of the decay-time fit to \Dz and \Dzb candidates in \dzkk decays is shown in Fig.~\ref{fig:result_time_tagged_tau_KK}.
After applying the mistag correction, the resulting value of \agamma is
\begin{displaymath}
\agamma = (-5.9\pm5.9_{\rm stat}\pm2.1_{\rm syst})\times 10^{-3}.
\label{eqn:agammaresult}
\end{displaymath}
Both results on \ycp and \agamma are compatible with zero and in agreement with previous measurements~\cite{Staric:2007dt,Aubert:2009ck,Aubert:2007en}.
Future updates are expected to lead to significant improvements in the sensitivity.
The systematic uncertainty is expected to be reduced by an improved treatment of background events which will be possible for the data taken in 2011.

\section*{Acknowledgements}

\noindent We express our gratitude to our colleagues in the CERN accelerator
departments for the excellent performance of the LHC. We thank the
technical and administrative staff at CERN and at the LHCb institutes,
and acknowledge support from the National Agencies: CAPES, CNPq,
FAPERJ and FINEP (Brazil); CERN; NSFC (China); CNRS/IN2P3 (France);
BMBF, DFG, HGF and MPG (Germany); SFI (Ireland); INFN (Italy); FOM and
NWO (The Netherlands); SCSR (Poland); ANCS (Romania); MinES of Russia and
Rosatom (Russia); MICINN, XuntaGal and GENCAT (Spain); SNSF and SER
(Switzerland); NAS Ukraine (Ukraine); STFC (United Kingdom); NSF
(USA). We also acknowledge the support received from the ERC under FP7
and the Region Auvergne.


\bibliographystyle{LHCb}
\bibliography{main_preprint}

\end{document}